\documentclass[american,aps,longbibliograph,reprint]{revtex4-2}
\usepackage[T1]{fontenc}
\usepackage[utf8]{inputenc}
\usepackage{amsmath}
\usepackage{graphicx}
\makeatletter
\usepackage[unicode=true]{hyperref}
\hypersetup{colorlinks}
\makeatother
\usepackage{babel}

\begin{document}
    \title{From elastic to inelastic deformation of a dipolar supersolid}
    \author{Qiaomei Zhao}
    \address{School of Physics and Materials Science, Guangzhou University, 230
        Wai Huan Xi Road, Guangzhou Higher Education Mega Center, Guangzhou
        510006, China}
    \address{School of Physics, Henan Normal University, Xinxiang, Henan 453007,
        China}
    \author{Xingdong Zhao}
    \email{phyzhxd@gmail.com}
    
    \address{School of Physics, Henan Normal University, Xinxiang, Henan 453007,
        China}
    \author{Jieli Qin}
    \email{qinjieli@126.com; 104531@gzhu.edu.cn}
    
    \address{School of Physics and Materials Science, Guangzhou University, 230
        Wai Huan Xi Road, Guangzhou Higher Education Mega Center, Guangzhou
        510006, China}
    \begin{abstract}
        Due to its peculiar superfluid-crystal duality feature, supersolid
        has received great research interest. Recently, researchers have paid
        much attention to its elastic response properties; however, the inelastic
        deformation has barely been explored. In this work, we study the transition
        from elastic to inelastic deformation of a dipolar supersolid Bose-Einstein
        condensate trapped in a box potential (i.e., a dipolar supersolid
        with finite size). We obtained the stationary supersolid states (both
        ground and excited) of the system, and examined the relation between
        the supersolid size and the number of unit cells it can accommodate,
        which can essentially help us to understand the dynamical responses
        of the supersolid during a dilation or compression process. We found
        that within a certain dilation or compression extent, the supersolid
        can retain its original crystal structure, that is, it endures an
        elastic deformation; however, when the extent exceeds a critical threshold,
        the original crystal structure of the supersolid will be disrupted,
        which signifies an inelastic deformation. Furthermore, we both analytically
        and numerically determined the critical point of the transition from
        elastic to inelastic deformation, and mapped out a phase diagram.
        These results open up new territory in the research of supersolid
        mechanical properties, and may find applications in quantum material
        science and quantum-based technologies.
    \end{abstract}
    \maketitle
    
    \section{Introduction}\label{sec:Introduction}
    
    Supersolid refers to the phenomenon that a quantum state of matter
    simultaneously shows the properties of superfluid and periodic crystal
    structure \citep{boninsegni_colloquium_2012}, the concept of which
    dates back to times as early as the middle of the 20th century, when
    researchers devoted great efforts to study the quantum properties
    of liquid Helium system \citep{gross_unified_1957,thouless_flow_1969,andreev_quantum_1969,chester_speculations_1970,leggett_can_1970};
    however, until nowadays no deterministic signature of supersolid Helium
    has been clearly experimentally observed \citep{chan_overview_2013,kim_probable_2004,kim_absence_2012,nyeki_intertwined_2017}.
    In recent years, much effort in supersolid studies has shifted to
    the ultracold atomic systems, and the supersolid states have already
    been successfully realized in several different platforms, including
    spin-orbit coupled Bose-Einstein condensates (BECs) \citep{li_stripe_2017,bersano_experimental_2019,putra_spatial_2020},
    cavity-BEC coupling systems \citep{leonard_supersolid_2017,leonard_monitoring_2017,schuster_supersolid_2020}
    and dipolar BECs \citep{tanzi_observation_2019,tanzi_supersolid_2019,guo_low-energy_2019,bottcher_transient_2019,chomaz_long-lived_2019,norcia_two-dimensional_2021,sohmen_birth_2021,bottcher_new_2020,recati_supersolidity_2023}.
    In the spin-orbit coupled BECs and cavity-BEC coupling systems, the
    supersolid crystal structure follows the externally applied laser
    lights, making it effectively rigid. While, in dipolar BECs, the supersolid
    crystal structure originates from the internal interactions between
    BEC atoms, allowing it to be compressed and stretched \citep{tanzi_supersolid_2019},
    thus making it an ideal platform to study the elastic and inelastic
    deformations of a supersolid during the dilation or compression process.
    
    When an ordinary solid material is subjected to a weak external force,
    the atoms are slightly displaced from their equilibrium positions
    leading to a small macroscopic deformation, nevertheless, the overall
    crystal structure of the sample remains intact, and the material can
    restore its original shape upon the removal of the external force.
    Such small deformations are typically considered to be elastic. Similar
    things also happen to the supersolids, and in refs. \citep{josserand_coexistence_2007,arpornthip_influence_2011,martone_supersolidity_2021,blakie_compressibility_2023,rakic_elastic_2024,poli_excitations_2024,platt_sound_2024,sindik_sound_2024,senarath_yapa_anomalous_2025},
    the elastic response properties of supersolids have been extensively
    examined. It is natural to further speculate that, similar to ordinary
    solid materials, imposing a large deformation on the supersolid would
    also disrupt its original crystal structure, resulting in a non-reversible
    inelastic deformation. However, we found such a hypothesis has not
    been verified yet, let alone the determination of the critical point
    of the transition from elastic to inelastic deformation.
    
    In this work, we investigate the elastic to inelastic deformation
    transition of a one-dimensional dipolar supersolid BEC confined within
    an external box potential well \citep{gaunt_bose-einstein_2013,navon_quantum_2021}.
    In many theoretical research works on supersolid, to identify the
    spontaneous breaking of spatial translation symmetry, researchers
    commonly consider an infinite large uniform system in the thermodynamic
    limit \citep{blakie_supersolidity_2020,martone_supersolidity_2021,blakie_compressibility_2023,rakic_elastic_2024,poli_excitations_2024,platt_sound_2024,smith_supersolidity_2023,ripley_two-dimensional_2023,kirkby_excitations_2024,sanchez-baena_superfluid-supersolid_2024,he_accessing_2025}.
    Here, we introduce the external potential to manipulate the dilation
    or compression process of the supersolid. Meanwhile, incorporating
    an external trap potential also makes the system more experimentally
    realistic. We choose the box trap, because it can best preserve the
    translational invariance, and brings as little disturbance as possible
    to the supersolid. We study the dilation and compression dynamics
    of the system in the context of an extended Gross-Pitaevskii theory,
    and we do observe a transition from elastic to inelastic deformation
    of the supersolid. We found this phenomenon can be well understood
    with the help of the stationary properties of the system, thus in
    this paper we also present the stationary supersolid states and their
    related properties, which allow us to construct an elastic to inelastic
    deformation phase diagram of the system both analytically and numerically.
    
    We organize the contents of this paper as follows. In Sec. \ref{sec:Model},
    we present the model studied in this work, and have a brief review
    on the roton instability and supersolid phenomenon of the system.
    In Sec. \ref{sec:Results}, we show the main results of this work.
    This section is further split into two sub-sections. In sub-Sec. \ref{subsec:StationarySupersolid},
    we show the stationary supersolid states and some properties of these
    states that will help us to understand the elastic and inelastic deformations
    of supersolids. In sub-Sec. \ref{subsec:Dynamics}, we show the transition
    from elastic to inelastic deformation of supersolids during a dilation
    or compression process, and explain this phenomenon using the stationary
    results obtained in the previous part. At last, we summarize this
    work in Sec. \ref{sec:summary}.
    
    \section{Model}\label{sec:Model}
    We consider a dipolar BEC subjected to a box potential in the $x$-direction,
    and very tightly confined by a harmonic trap in the transverse ($y$ and $z$) direction, 
    such that the transverse dynamics are frozen to the ground state. Such a system can be 
    reduced from three-dimensional to quasi-one-dimensional
    \citep{sinha_cold_2007,deuretzbacher_ground-state_2010,turmanov_generation_2020}, 
    and its dynamics are governed by the following dimensionless
    extended Gross-Pitaevskii equation (eGPE)
    \citep{turmanov_generation_2020,edmonds_quantum_2020,turmanov_oscillations_2021,gao_breathing_2021}
    \begin{align}
        i\frac{\partial\psi}{\partial t} & =\left[-\frac{1}{2}\frac{\partial^{2}}{\partial x^{2}}+V\left(x,t\right)+g\left|\psi\right|^{2}+\gamma\left|\psi\right|^{3}+p\left|\psi\right|^{4}\right]\psi\nonumber \\
        & +g_{dd}\psi\int R\left(\left|x-x'\right|\right)\left|\psi\left(x',t\right)\right|^{2}dx',\label{eq:eGPE}
    \end{align}
    where $V\left(x,t\right)=\frac{V_{0}}{2}\left\{ \tanh\left[\frac{x-w\left(t\right)}{\sigma}\right]-\tanh\left[\frac{x+w\left(t\right)}{\sigma}\right]+2\right\} $
    is the box potential with $V_{0}$, $w\left(t\right)$, $\sigma$
    characterizing its depth, time-dependent half-width (we will simply
    call it ``width'' in the following contents) and stiffness at edge;
    $g$ is the strength of short-range contact interaction; $\gamma$
    is the strength of effective Lee-Huang-Yang interaction which originates
    from the quantum fluctuations; $p$ is the strength of three-body
    interaction; $g_{dd}$ is the long-range dipole-dipole interaction
    strength,  $R\left(x\right)=-2\left|x\right|+\sqrt{\pi}\exp\left(x^{2}\right)\left(1+2x^{2}\right)\mathrm{erfc}\left(\left|x\right|\right)$
    is the one-dimensional dipole-dipole interaction response function
    \citep{sinha_cold_2007,deuretzbacher_ground-state_2010}, and the
    BEC wavefunction $\psi$ is normalized to the total atom number
    $N=\int\left|\psi\right|^{2}dx$.
    The dependence of dimensionless parameters $g, \gamma, p, g_{dd}$ on real physical
    quantities
    and their experimentally relevant values can be found in refs.
    \citep{turmanov_generation_2020,turmanov_oscillations_2021} (especially 
    ref. \citep{turmanov_oscillations_2021}, which also studies the dipolar 
    supersolid). In the present work, 
    we choose their values roughly the same as in these references.
    
    In such a system, the BEC naturally behaves like a superfluid, and
    a delicate balance between the short-range contact interaction and
    long-range dipole-dipole interaction leads to a spatial periodic modulation
    of the atomic density, the second ingredient of a supersolid. But,
    the system suffers a collapse under the sole existence of contact
    and dipole-dipole interactions, so to maintain a stable supersolid,
    it is essential to include the quantum fluctuations described by the
    Lee-Huang-Yang term, which can prevent the system from collapse
    \citep{bottcher_new_2020,recati_supersolidity_2023}.
    It has been shown that the elastic three-body interaction plays a
    similar role in stabilizing the dipolar BECs \citep{gautam_self-trapped_2019,turmanov_oscillations_2021,bisset_crystallization_2015,xi_droplet_2016},
    therefore in this work we also include it in the model.
    
    \begin{figure}
        \begin{centering}
            \includegraphics{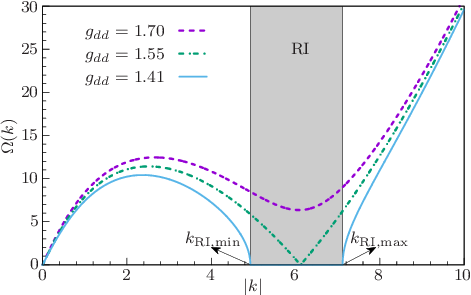}
            \par\end{centering}
        \caption{Collective excitation spectra of uniform dipolar BECs with different
            dipole-dipole interaction strength, Eq. (\ref{eq:collectiveExcitationSpectrum}).
            When $g_{dd}=1.70$, the curve of $\Omega\left(k\right)$ shows a
            roton minimum (dashed violet line). Around $g_{dd}=1.55$, the roton
            minimum drops to zero (dash-dotted green line). When $g_{dd}=1.41$,
            $\Omega\left(k\right)$ becomes complex valued in the gray color region
            $\left|k\right|\in\left[k_{\mathrm{RI,min}},k_{\mathrm{RI,max}}\right]$,
            which means a roton instability (RI, solid cyan line). Other parameters
            are $\rho_{0}=40$, $g=-1$, $\gamma=0.001$ and $p=0.006$. }\label{fig:CES}
    \end{figure}
    
    Without the external potential, $V\left(x,t\right)=0$, the system
    is translationally invariant, and it admits a uniform solution $\psi_{0}=\sqrt{\rho_{0}}\exp\left(-i\mu t\right)$
    with $\rho_{0}$ being the atomic density, and $\mu=g\rho_{0}+2g_{dd}\rho_{0}+\gamma\rho_{0}^{3/2}+p\rho_{0}^{2}$
    being the chemical potential. Under an appropriate range of parameters,
    this uniform solution is modulation unstable. To show the instability,
    we introduce a small perturbation on the uniform background, i.e.,
    we let $\psi=\left[\sqrt{\rho_{0}}+\delta\psi\left(x,t\right)\right]\exp\left(-i\mu t\right)$
    (with $\delta\psi\ll\sqrt{\rho_{0}}$). Further writing $\delta\psi\left(x,t\right)$
    into the Bogoliubov transformation form $\delta\psi=U\exp(ikx)\exp(-i\Omega t)+V\exp(-ikx)\exp(i\Omega^{*}t)$
    ($U,V$ are the perturbation amplitudes, and $k$ is the quasi momentum),
    and inserting it into Eq. (\ref{eq:eGPE}), one can obtain the following
    collective excitation spectrum of the system
    \begin{equation}
        \left|\frac{\Omega}{k}\right|=\sqrt{\frac{k^{2}}{4}+\rho_{0}\left[g+g_{dd}\tilde{R}\left(k\right)+\frac{3}{2}\gamma\rho_{0}^{1/2}+2p\rho_{0}\right]},\label{eq:collectiveExcitationSpectrum}
    \end{equation}
    where $\tilde{R}\left(k\right)=2\left[1+k^{2}\exp\left(k^{2}/4\right)\mathrm{Ei}\left(-k^{2}/4\right)/4\right]$
    is the Fourier transform of dipole-dipole interaction response function
    $R\left(x\right)$, with $\mathrm{Ei}\left(\cdot\right)$ being the
    exponential integral function. As shown in Fig. \ref{fig:CES}, the
    collective excitation spectrum exhibits a phonon-maxon-roton structure,
    and as $g_{dd}$ drops below a critical value, $\Omega$ takes complex
    value in the roton region, signifying a roton instability of the dipolar
    gas. As the roton instability happens, the matterwaves with wave number
    in the roton instability region, $\left|k\right|\in\left[k_{\mathrm{RI,min}},k_{\mathrm{RI,max}}\right]$,
    will be automatically amplified because of the exponential growth
    feature of factors $\exp(-i\Omega t)$ or $\exp(i\Omega^{*}t)$ when
    $\Omega$ is a complex number, and the interference of $\exp(\pm ikx)$
    waves will lead to the spontaneous formation of a stripe pattern,
    i.e., the supersolid phenomenon.
    
    \section{Results}\label{sec:Results}
    
    In this work, we introduce a deep box potential (which can best preserve
    the spatial uniform property) to truncate the supersolids to finite
    size, so that we can study their elastic and inelastic deformations
    during the dilation and compression processes by dynamically quenching
    the width of the box potential. We found that the dynamics in the
    dilation and compression processes can be well understood with the
    help of the stationary properties of the supersolid states. Thus,
    in this section, we will first show the stationary supersolid states
    and the properties related to their elastic and inelastic deformations
    in sub-Sec. \ref{subsec:StationarySupersolid}; and put the results
    of dilation and compression dynamics afterward in sub-Sec. \ref{subsec:Dynamics}. 
    
    \subsection{Stationary supersolid states}\label{subsec:StationarySupersolid}
    
    \begin{figure}
        \begin{centering}
            \includegraphics{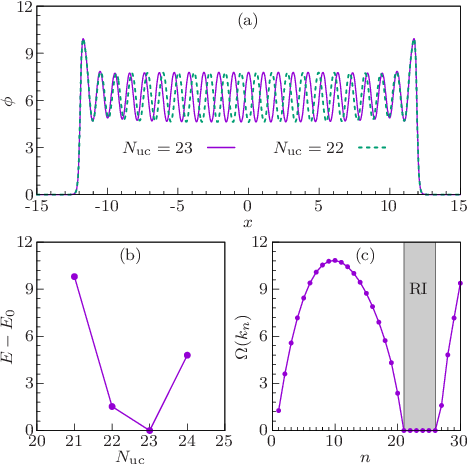}
            \par\end{centering}
        \caption{Stationary supersolid states in a box potential with width $w=12$.
            (a) The ground state which contains $N_{uc}=23$ unit cells (solid
            violet line), and the excited state which contains $N_{uc}=22$ unit
            cells (dashed green line). (b) The stationary energy as a function
            of unit cell numbers, with the zero energy point shifted to the ground
            state energy. 
            (c) The discrete collective excitation spectrum
            $\Omega\left(k_{n}\right)$ with $k_{n}=\left(n-1/2\right)\pi/w$
            and $\rho_0 = N/\left(2w\right)$.
            The roton instability (RI) region is highlighted in gray color. In
            panels (b,c), the solid line just connects the data points for guiding
            the eye. The parameters used are $N=1000$, $g=-1$, $g_{dd}=1.41$,
            $\gamma=0.001$, $p=0.006$, $V_{0}=100$, and $\sigma=0.1$ (these
            parameters will remain unchanged in the following contents).}\label{fig:supersolidWidth12.0}
    \end{figure}
    
    We assume a time-independent box potential {[}i.e., $w\left(t\right)=w${]},
    and solve the stationary state problem of the system. We let $\psi\left(x,t\right)=\phi\left(x\right)\exp\left(-i\mu t\right)$
    with $\mu$ being the chemical potential, and insert it into Eq. (\ref{eq:eGPE}),
    we obtain a stationary eGPE
    \begin{align}
        \mu\phi & =\left[-\frac{1}{2}\frac{\partial^{2}}{\partial x^{2}}+V\left(x\right)+g\left|\phi\right|^{2}+\gamma\left|\phi\right|^{3}+p\left|\phi\right|^{4}\right]\phi\nonumber \\
        & +g_{dd}\phi\int R\left(\left|x-x'\right|\right)\left|\phi\left(x'\right)\right|^{2}dx'.\label{eq:stationaryeGPE}
    \end{align}
    Solving this nonlinear eigenvalue problem, we obtain the stationary
    states of the system. Due to the complexity of the interactions and
    trapping potential,
    it is impossible to solve this nonlinear integro-differential equation analytically, 
    therefore we discretize it (typically with 2048 points in the range of $x\in[-24,24]$, 
    which is 4/3 times the maximum trap width considered in this work), and solve the 
    discretized problem numerically with a trust-region-dogleg nonlinear solver
    \citep{powell_AFortran_1968,More_User_1980} with accuracy tolerance typically set to $10^{-6}$.
    
    In Fig. \ref{fig:supersolidWidth12.0}, we showcase the stationary
    supersolid states in a box potential with width $w=12.$ We found
    this box potential can support four different stationary supersolid
    states that contain $N_{uc}=$21, 22, 23 and 24 unit cells respectively.
    Among them, the supersolid state with 23 unit cells has the lowest
    energy, thus is the ground state, and the others are excited states.
    The energies of these states are plotted in panel (b). In panel (a),
    we show the wavefunctions of the ground and the first excited supersolid
    state. In the bulk region, we clearly observe both a periodic crystal
    structure and a superfluid background, which are the main features
    of a supersolid state. Near the edges of the box, we observe two peaks,
    which are apparently higher than the supersolid peaks in the bulk
    region. They result from the edge effect of a trapped dipolar gas
    \citep{roccuzzo_supersolid_2022,juhasz_how_2022}. 
    
    Here, for the box width $w=12$, we can not find supersolid states
    containing unit cells more than 24 or less than 21, i.e., for a fixed
    box width, there exists both a maximum and a minimum number of unit
    cells. This can be understood as follows. The roton instability
    in region $\left|k\right|\in\left[k_{\mathrm{RI,min}},k_{\mathrm{RI,max}}\right]$
    leads to the formation of supersolid states.  When a deep box
    potential with width $w$ is imposed on the dipolar BEC, it acts like
    a Fabry-Perot type mode selector, and selects only the modes
    $\phi_{m}=\sin\left[k_{m}\left(x+w\right)\right]$
    with discrete wavenumbers $k_{m}=m\pi/\left(2w\right)$, $m=1,2,3,4,\cdots$.
    Furthermore, since the modes with $m=2,4,\cdots$ have the odd parity,
    they can not co-exist with the constant superfluid background which
    has the even parity. Therefore, a finite-size supersolid trapped in
    a box potential with width $w$ should fulfill the relationship
    $k_{\mathrm{RI,min}}<k_{n}=\left(n-1/2\right)\pi/w<k_{\mathrm{RI,max}}$
    with $n=1,2,\cdots$. Here, $n$ determines the number of maximums
    of the mode function, therefore it can be interpreted as the number
    of unit cells contained in the supersolid, i.e., $n=N_{\mathrm{uc}}$.
    In such a way, for a fixed box width, the number of unit cells contained
    in the supersolid is bounded from both above and below. 
    
    \begin{figure}
        \begin{centering}
            \includegraphics{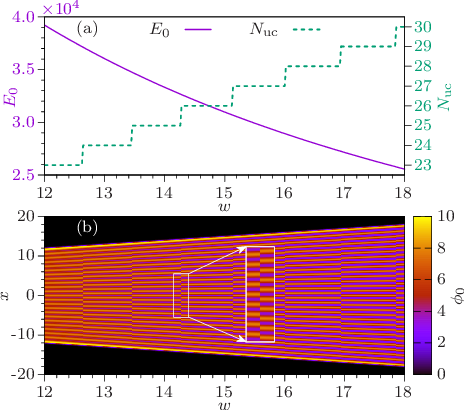}
            \par\end{centering}
        \caption{Ground supersolid states in box potential as the width changes. (a)
            Ground state energy $E_{0}$ (left axis, solid violet line) and unit
            cell number $N_{\mathrm{uc}}$(right axis, dashed green line) as a
            function of the box width $w$. (b) The ground supersolid state $\phi_{0}\left(x\right)$
            as a function of the box width $w$. The inset enlarges the small
            boxed area, to clearly show the abrupt change of the wavefunction.
        }\label{fig:GS}
    \end{figure}
    
    \begin{figure}
        \begin{centering}
            \includegraphics{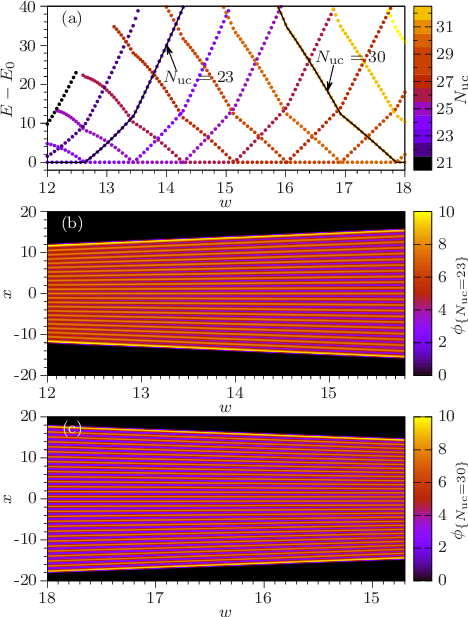}
            \par\end{centering}
        \caption{Excited supersolid states in box potential as the width changes. (a)
            Energy spectrum. The lowest few $E_{i}-E_{0}$ values are plotted
            as a function of the box width $w$. The color of these points represents
            the number of unit cells contained in the corresponding supersolid
            state. The points for $N_{\mathrm{uc}}=23$ and $30$ are connected
            with a black line for guiding the eye. (b,c) Wavefunctions of the
            supersolid state with $N_{\mathrm{uc}}=23$ (b) and $30$ (c) unit
            cells, as a function of box width $w$. These two panels terminate
            at $w=15.8$ and $14.7$ respectively, because when $w>15.8$, the
            supersolid state with $N_{\mathrm{uc}}=23$ no longer exists; and
            when $w<14.7$, the supersolid state with $N_{\mathrm{uc}}=30$ also
            does not exist any more. }\label{fig:ES}
    \end{figure}
    
    In panel (c) of Fig. \ref{fig:supersolidWidth12.0}, we plot the discrete
    collective excitation spectrum $\Omega\left(k_{n}\right)$ for uniform atomic density 
    $\rho_0 =N/\left(2w\right)$ with $w=12$. 
    It is seen that the roton instability predicts that there can be $N_{\mathrm{uc}}=21\sim26$
    unit cells in the supersolid,
    which roughly agrees with our numerical results. The discrepancy
    may originate from the following two reasons. First, in the roton instability analysis, 
    we assume a uniform atomic density; 
    however, the box trapped system is only approximately uniform 
    in the central region.
    Secondly, the roton instability can only approximately signify 
    (instead of precisely determine) the occurrence of supersolid phenomenon. 
    In fact, in previous studies \citep{blakie_supersolidity_2020,smith_supersolidity_2023,ripley_two-dimensional_2023},
    it has been shown that the roton instability can only approximately
    determine the boundary of the supersolid phase. 
    
    We further examined the ground supersolid states when the box width
    changes, with the main results shown in Fig. \ref{fig:GS}. As shown
    by the violet solid line in panel (a), when the box width $w$ increases,
    the atomic density is reduced, which weakens the interactions between
    BEC atoms, thus leads to the gradual decrease of the ground state energy
    $E_{0}$. We also observe that when the trapping box becomes wider,
    the ground supersolid state can contain more and more unit cells,
    see the green dashed line in panel (a), which plots the relation between
    $N_{\mathrm{uc}}$ and $w$, and it jumps step by step as $w$ increases.
    Every time when $N_{\mathrm{uc}}$ jumps, the ground state supersolid
    wavefunction also endures an abrupt change, see the panel (b) and
    the inset in it. 
    
    The results for the excited supersolid states are shown in Fig. \ref{fig:ES}.
    In panel (a), we plot the stationary supersolid state energies (with
    the ground state energy being subtracted) as a function of trapping
    box width, with the color of the point indicating the number of unit
    cells contained in the supersolid. We found that the sets of points
    with the same number
    of unit cells are connecting to a continuous curve, for example, we explicitly
    connect the points of $N_{\mathrm{uc}}=23$ and $30$
    as an illustration. We also plot the stationary supersolid wavefunctions
    with the same number of unit cells when the trapping box width changes,
    see panels (b) and (c) for $N_{\mathrm{uc}}=23$ and $30$.
    It can be seen that the wavefunctions $\phi_{\left\{ N_{\mathrm{uc}}=23\right\} }$
    and $\phi_{\left\{ N_{\mathrm{uc}}=30\right\} }$ change smoothly
    when the trapping box width $w$ increases or decreases.
    Compared to the abrupt changes of the ground state wavefunction, we suspect that 
    when the box potential is dynamically stretched or compressed, 
    the ground supersolid state can not follow up correspondingly;
    instead, the system will rather follow these excited stationary states. 
    In section \ref{subsec:Dynamics}, we will verify this by numerically
    simulating the time-dependent eGPE (\ref{eq:eGPE}).
    
    \begin{figure}
        \begin{centering}
            \includegraphics{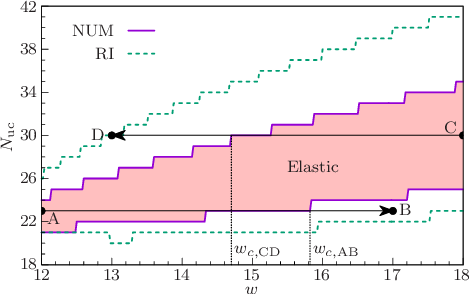}
            \par\end{centering}
        \caption{Elastic and inelastic deformation phase diagram of a finite-size supersolid.
            The maximum and minimum number of unit cells of a supersolid trapped in a box well 
            with width $w$ are plotted. 
            The two violet solid lines are the numerical (NUM) results, while the two
            green dashed lines are the results obtained by analytical roton instability
            analysis (RI). The pink area can be seen as the elastic deformation
            region, while quenching out this region
            will lead to an inelastic deformation. The two arrows AB and CD indicate
            two quenching processes, with $w_{c,\mathrm{AB}}=15.8$ and $w_{c,\mathrm{CD}}=14.7$
            being the corresponding elastic to inelastic transition critical points.  }\label{fig:PhaseDiagram}
    \end{figure}
    
    Since the number of unit cells contained in the supersolid is bounded
    from both above and below, we show a figure plotting the maximum and
    minimum number of the unit cells contained in a box-trapped supersolid,
    see Fig. \ref{fig:PhaseDiagram}, where the violet solid lines represent
    the numerical result, and the green dashed lines are obtained by the
    roton instability analysis. It is seen that they qualitatively agree
    with each other
    [the discrepancy comes from the same reasons as in 
    Fig. \ref{fig:supersolidWidth12.0}(c)].
    This figure is the most significant result of the present
    paper, which would be regarded as an \emph{elastic and inelastic deformation
        phase diagram} of the dipolar supersolids. In the next part, we will
    show that it can essentially help us to understand the deformation
    dynamics of the supersolids during the dilation and compression processes.

    \subsection{Transition from elastic to inelastic deformation}\label{subsec:Dynamics}
    
    We initially prepare the system in the ground supersolid state of
    a box potential well with width $w_{i}$, then we dynamically quench
    the width of the box to explore the dilation and compression
    dynamics of the supersolid. Here, we adopt a linear quenching scheme,
    that is the width of the box depends on time $t$ as follows
    \begin{equation}
        w\left(t\right)=w_{i}+\frac{w_{f}-w_{i}}{\tau_{Q}}t,\label{eq:quenchingScheme}
    \end{equation}
    where $w_{f}$ is the destination box width, and $\tau_{Q}$
    is the quenching time.
    We numerically solve the time-dependent eGPE (\ref{eq:eGPE}) using the split-step
    Fourier method with the spatial discretizing parameters the same as in the 
    stationary states calculation, and the time step is typically set to $dt=10^{-4}$.
    
    \begin{figure}
        \begin{centering}
            \includegraphics{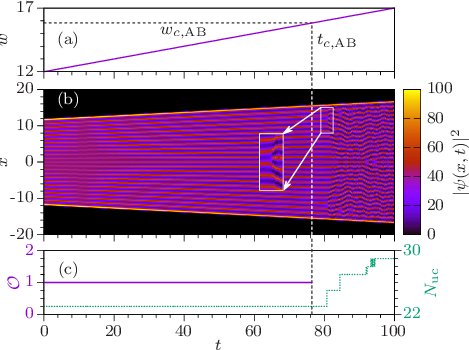}
            \par\end{centering}
        \caption{Transition from elastic to inelastic deformation of a supersolid during a dilation
            quenching process indicated by the arrow AB in Fig. \ref{fig:PhaseDiagram}.
            (a) The quenching scheme, i.e., Eq. (\ref{eq:quenchingScheme})
            with $w_{i}=12$, $w_{f}=17$ and $\tau_{Q}=100$. (b) The time evolution
            of atomic density $\left|\psi\left(x,t\right)\right|^{2}$. The inset
            enlarges the small boxed area to clearly show the bifurcating of the
            supersolid sites. (c) 
            The overlap $\mathcal{O}\left(t\right)$ between $\psi$ and 
            $\phi_{N_{\mathrm{uc}}=23}$ (violet solid line), 
            and number of supersolid unit cells (green dashed line) contained
            in the box trapped supersolid during this dilation process.
        }\label{fig:dilation}
    \end{figure}
    
    \begin{figure}
        \begin{centering}
            \includegraphics{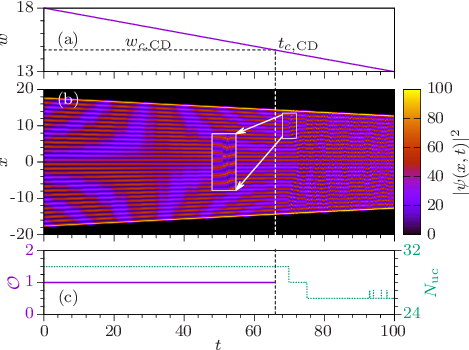}
            \par\end{centering}
        \caption{Transition from elastic to inelastic deformation of a supersolid during a compression
            quenching process indicated by the arrow CD in Fig. \ref{fig:PhaseDiagram}. 
            (a) The quenching scheme, i.e., Eq. (\ref{eq:quenchingScheme})
            with $w_{i}=18$, $w_{f}=13$ and $\tau_{Q}=100$. (b) The time evolution
            of atomic density $\left|\psi\left(x,t\right)\right|^{2}$. The inset
            enlarges the small boxed area to clearly show the merging of the supersolid
            sites. (c) 
            The overlap $\mathcal{O}\left(t\right)$ between $\psi$ and
            $\phi_{N_{\mathrm{uc}}=30}$ (violet solid line), 
            and number of supersolid unit cells (green dashed line) contained
            in the box trapped supersolid during this compression process.
        }\label{fig:compression}
    \end{figure}
    
    Firstly, we come to consider the concrete dilation quenching process
    indicated by the arrow AB in Fig. \ref{fig:PhaseDiagram}. Initially,
    we set the box width to $w_{i}=12$, and the initial ground supersolid
    state contains $N_{\mathrm{uc}}=23$ unit cells. When we dynamically
    increase the box width in the range $w<w_{c,\mathrm{AB}}$, it can
    be imagined that the system will not always remain in the ground state,
    because the ground state can not smoothly follow the broadening
    of the box potential (we have shown this in Fig. \ref{fig:GS} );
    instead, we would expect that the system will stay in the $\phi_{\left\{ N_{\mathrm{uc}}=23\right\} }$
    state, because it is this state that can smoothly follow the broadening
    of the box poential (this has been shown in Fig. \ref{fig:ES}). This
    means that in this stage, the global crystal structure of the supersolid
    remains unchanged, only the size of the unit cells grows larger to some extent. We
    can recognize this type of deformation of the supersolid as an elastic
    one. Once the box width $w$ exceeds the critical value $w_{c,\mathrm{AB}}$,
    the stationary state $\phi_{\left\{ N_{\mathrm{uc}}=23\right\} }$
    becomes inaccessible, this will force a change of the unit cell number $N_{\mathrm{uc}}$
    of the supersolid, i.e., a change of its global crystal structure.
    We would recognize this type of deformation as an inelastic one. Similar
    things would also happen in the compression process indicated by the arrow
    CD in Fig. \ref{fig:PhaseDiagram}. 
    We expect that before the box width $w$ decreases to the critical
    value $w_{c,\mathrm{CD}}$, the deformation is elastic, and surpassing
    this critical value leads to an inelastic deformation of the supersolid
    as well. In such a way, in Fig. \ref{fig:PhaseDiagram}, the lines
    of maximum and minimum value of $N_{\mathrm{uc}}$ can be taken as
    the boundary between elastic and inelastic deformation,
    with the in-between pink color region being the elastic region. 
    
    To verify the above scenario, we simulate the dilation and compression
    quenching dynamics by numerically solving the eGPE (\ref{eq:eGPE}),
    with the results shown in Figs. \ref{fig:dilation} and \ref{fig:compression}.
    In these two figures, before the critical points, we do observe a
    smooth spatial expanding or shrinking of the atomic density profile with
    the supersolid crystal structure being intact.
    We also examined the overlap between the dynamical wavefunction
    and the corresponding stationary excited states, i.e., 
    \begin{equation}
        \mathcal{O}(t) = \left|
        \int \phi_{N_\mathrm{uc}=23,30}\left(x,w\left(t\right)\right)\psi^{*}
        \left(x,t\right)dx
        \right|/N. \label{eq:overlap}
    \end{equation}
    As shown by the violet solid line in panels (c), the
    overlap remains nearly 1, which means the dynamical wavefunction
    $\psi(x,t)$ follows the excited stationary states $\phi_{N_\mathrm{uc}=23,30}$ 
    very well.
    Shortly after the box width $w$ surpasses the critical points
    $w_{c,\mathrm{AB}}$ and $w_{c,\mathrm{CD}}$
    predicted by Fig. \ref{fig:PhaseDiagram},
    we observe events of supersolid sites
    bifurcating or merging, which destroy the original
    supersolid crystal structure.
    This resembles the
    stimulation of dislocations \citep{arpornthip_influence_2011}. After
    the bifurcating or merging happens, severe collective excitations
    are aroused \citep{turmanov_oscillations_2021}. All these observations
    illustrate a transition from elastic to inelastic deformation,
    agreeing well with the scenario discussed in the previous
    paragraph;
    therefore, we conclude that Fig. \ref{fig:PhaseDiagram} can be regarded
    as an elastic and inelastic deformation phase diagram.
    
    We also note that a change of supersolid unit cell number also happens
    during the superfluid to supersolid phase transition \citep{kirkby_kibble-zurek_2025}, 
    but in this process, new unit cells emerge from the flat superfluid background; 
    no bifurcation and merging of unit cells have been observed. Moreover, in our work,
    the parameters always fall in the supersolid phase; 
    thus, no superfluid to supersolid phase transition happens in our system, 
    and the bifurcation and merging of supersolid unit cells are the mechanical 
    response to the stretching and compressing of the system, 
    i.e., an inelastic deformation.

    \section{Summary}\label{sec:summary}
    
    In summary, we show that similar to ordinary solid materials, during
    a dilation or compression process, a dipolar supersolid also experiences
    an elastic deformation within a certain extent; and exceeding this
    extent, the deformation becomes inelastic. We found that this phenomenon
    can be well understood using the stationary properties of the supersolid
    states, based on which we have mapped out an elastic to inelastic
    phase transition diagram of the system using both numerical and analytical
    approaches. Our work pushes the research of supersolid mechanical
    properties from the elastic region to the inelastic region. Since
    the dilation and compression processes of BECs are useful in building
    quantum heat engines
    \citep{li_efficient_2018,myers_boosting_2022,li_quantum_2022,simmons_thermodynamic_2023,estrada_quantum_2024},
    the results of this work would also be meaningful in the fields of
    quantum thermodynamics and quantum-based technologies,
    and these might be interesting future research subjects.
    
    \begin{acknowledgments}
        This work is supported by Guangdong Basic and Applied Basic Research
        Foundation (2024A1515012526), National Natural Science Foundation
        of China (11904063), and Natural Science Foundation of Henan‌ Province (242300420255). 
    \end{acknowledgments}
    
    %

\end{document}